%% file: pi0pi0hc_prl.tex
\RequirePackage{lineno}
\documentclass[twocolumn,showpacs,superscriptaddress,amsmath,amssymb]{revtex4-1}
\usepackage{graphicx}
\usepackage{epsfig}
\usepackage{dcolumn}
\usepackage{bm}
\usepackage{color}
\usepackage{lineno}
\usepackage{subfigure}
\include{def-com}

\def \zcp {Z_c(4020)}
\def \zcpn {Z_c(4020)^0}
\def \EE  {e^+e^-}
\def \pphc {\pi^{0}\pi^{0}h_{c}}

\def \jp {J/\psi}
\def \ppjpsi {\pi^+\pi^-\jp}

\def \EE {e^+e^-}

\def \kk {K^+K^-}
\def \pp {\pi^+\pi^-}
\def \ppp {\pp\pi^0}
\def \pp {\pi^+\pi^-}
\def \kk {K^+K^-}
\def \ppb {p\bar{p}}
\def \ppkk {\pi^+\pi^-K^+K^-}
\def \ppkk {\pi^+\pi^-K^+K^-}
\def \jpsi {J/\psi}
\begin{document}

\title{\boldmath
{Observation of $\EE\to \pphc$ and a neutral charmoniumlike structure} $\zcpn$ }

\author{
\begin{small}
\begin{center}
M.~Ablikim$^{1}$, M.~N.~Achasov$^{8,a}$, X.~C.~Ai$^{1}$,
      O.~Albayrak$^{4}$, M.~Albrecht$^{3}$, D.~J.~Ambrose$^{42}$,
      A.~Amoroso$^{46A,46C}$, F.~F.~An$^{1}$, Q.~An$^{43}$, J.~Z.~Bai$^{1}$
% ...
      \\
      \vspace{0.2cm}
      (BESIII Collaboration)\\
      \vspace{0.2cm} {\it
        $^{1}$ Institute of High Energy Physics, Beijing 100049, People's Republic of China\\
% ...
$^{3}$ Bochum Ruhr-University, D-44780 Bochum, Germany\\
        $^{4}$ Carnegie Mellon University, Pittsburgh, Pennsylvania 15213, USA\\
% ...
$^{8}$ G.I. Budker Institute of Nuclear Physics SB RAS (BINP), Novosibirsk 630090, Russia\\
% ...
$^{42}$ University of Rochester, Rochester, New York 14627, USA\\
        $^{43}$ University of Science and Technology of China, Hefei 230026, People's Republic of China\\
% ...
$^{46}$ (A)University of Turin, I-10125, Turin, Italy; (B)University of Eastern Piedmont, I-15121, Alessandria, Italy; (C)INFN, I-10125, Turin, Italy\\
% ...
$^{a}$ Also at the Novosibirsk State University, Novosibirsk, 630090, Russia\\
% ...
~\\~\\
      }\end{center}
    \vspace{0.4cm}
  \end{small}
}

\begin{abstract}
Using data collected with the BESIII detector operating at the Beijing
Electron Positron Collider at center-of-mass energies of
$\sqrt{s}=4.23$, 4.26, and 4.36~GeV, we observe $\EE\to \pphc$ for the first time.  The
Born cross sections are measured and found to be about half of those
of $\EE\to \pi^+\pi^-h_c$ within less than 2$\sigma$. In the $\pi^0h_c$ mass spectrum,
a structure at 4.02~GeV/$c^2$ is found. It is most likely to be the neutral isospin
partner of the $\zcp^{\pm}$ observed in
the process of $\EE\to \pi^+\pi^-h_c$ is found.
A fit to the $\pi^0 h_c$ 
invariant mass spectrum, with the width
of the $\zcpn$ fixed to that of its charged isospin partner
and possible interferences with non-$\zcpn$ amplitudes neglected, gives
a mass of ($4023.9\pm 2.2 \pm 3.8$)~MeV/$c^2$ for the
$\zcpn$, where the first error is statistical and the second
systematic.
\end{abstract}


\maketitle

In the study of $\EE\to \ppjpsi$, a distinct charged structure,
$Z_c(3900)^{\pm}$, was observed in the $\pi^\pm \jpsi$ spectrum by the
BESIII~\cite{ref1} and Belle~\cite{ref2} experiments, and confirmed
shortly thereafter with CLEO-c data~\cite{CLEO}. A similar charged
structure but with a slightly higher mass, $Z_{c}(4020)^{\pm}$, was soon reported
in $\EE\to \pp h_c$~\cite{guoyp} by BESIII.  As
there are at least four quarks within these two charmoniumlike structures, they
are interpreted as either tetraquark states, $D\bar{D}^*$ (or $D^*\bar{D}^*$) molecules,
hadrocharmonia, or other configurations~\cite{ref4}. More recently,
charged structures in the same mass region were observed via their
decays into charmed meson pairs, including the charged $Z_c(4025)^{\pm}$ in $\EE\to
\pi^{\pm} (D^*\bar{D}^*)^{\mp}$~\cite{ref5} and the charged $Z_c(3885)^{\pm}$ in $\EE\to
\pi^{\pm} (D\bar{D}^*)^{\mp}$~\cite{ref5add}. These structures together
with the recently confirmed $Z(4430)^-$~\cite{belle4430,belle4430conf,lhcb4430} and
similar structures observed in the bottomonium system~\cite{ref6} indicate
that a new class of hadrons has been observed.
An important question is whether
all these charged structures are part of isospin $I=1$
triplets, in which case neutral partners with $I_z = 0$ should also
be found.
Evidence for a neutral $Z_c(3900)$ was observed in $\EE\to \pi^0\pi^0 J/\psi$ process with CLEO-c data
at center-of-mass energy (CME) $\sqrt{s}$=4.17~GeV~\cite{CLEO}. A neutral structure, the $Z_c(4020)^{0}$,
is expected to couple to the $\pi^0 h_c$ final state and be produced for in $\EE\to \pphc$ processes.

In this Letter, we present the first observation of $\EE\to \pphc$
at $\sqrt{s} = 4.23$~GeV, 4.26~GeV, and 4.36~GeV, and
the observation of a neutral charmoniumlike structure $\zcpn$ in the $\pi^0h_c$
spectrum. We closely follow the analysis of $\EE\to \pp
h_c$~\cite{guoyp} with the selection of $\pp$ replaced with the
selection of a pair of $\pi^{0}$s. The data samples were collected
with the BESIII detector~\cite{bepc}. The CMEs and corresponding integrated luminosities are listed in
Table~\ref{data}.

  We use a GEANT4 based~\cite{geant4} Monte Carlo (MC) simulation to
  optimize the event selection
  criteria, determine the detection efficiency, and estimate
  backgrounds.
In the studies presented here, the $h_c$ is
reconstructed via its electric-dipole (E1) transition $h_c\to
\gamma\eta_c$ with $\eta_c\to X_i$, where $X_i$ denotes 16
hadronic final states: $\ppb$, $\ppkk$, $\pp\ppb$, $2(\kk)$, $2(\pp)$,
$3(\pp)$, $2(\pp)\kk$, $K_SK^{\pm}\pi^{\mp}$,
$K_SK^{\pm}\pi^{\mp}\pp$, $\kk\pi^0$, $\kk\eta$, $\ppb\pi^0$, $\pp\eta$,
$\pp\pi^0\pi^0$, $2(\pp)\eta$, and $2(\ppp)$.
The initial state radiation (ISR) is simulated with KKMC~\cite{KKMC},
where the Born cross section of $\EE\to \pphc$ is assumed to follow
the $\EE\to \pp h_c$ line-shape~\cite{guoyp}.

\begin{table*}[htbp]
\caption{Energies ($\sqrt{s}$), luminosities ($\cal L$),
numbers of events ($n_{h_c}^{\rm obs}$), average efficiencies
($\sum\limits^{16}_{i=1}\epsilon_i{\cal B}(\eta_c\ar X_i)$), initial
state radiative correction factor ($1+\delta^{r}$)~\cite{guoyp},
vacuum polarization factor ($1+\delta^v$),
Born cross sections $\sigma^{\rm B}(\EE\to \pphc)$ and ratios ${\cal R}_{\pi\pi h_c}
= \frac{\sigma(\EE\to \pphc)} {\sigma(\EE\to \pp h_c)}$, where the third errors
are from the uncertainty in ${\cal B}(h_c\to
\gamma\eta_c)$~\cite{ref11}.}  \bcl \doublerulesep 2pt
\begin{tabular}{cccccccc}\hline\hline
$\sqrt{s}~(\rm GeV)$ & ${\cal L}~({\rm pb}^{-1})$ & $n_{h_c}^{\rm
obs}$ &$\sum\limits^{16}_{i=1}\epsilon_i{\cal B}(\eta_c\ar X_i)$ &$1+\delta^r$&$1+\delta^v$ &$\sigma^{\rm B}(\EE\to \pphc)~(\rm pb)$ & ${\cal R}_{\pi\pi h_c}$\\
\hline
4.230&1090.0&$82.5\pm15.6$&$6.82\times10^{-3}$&0.756&1.056&$25.6\pm4.8\pm2.6\pm4.0$&$0.54\pm0.11\pm0.06$\\
4.260&826.8&$62.8\pm13.3$&$6.54\times10^{-3}$&0.831&1.054&$24.4\pm5.2\pm3.2\pm3.8$&$0.63\pm0.14\pm0.10$\\
4.360&544.5&$64.3\pm11.5$&$6.68\times10^{-3}$&0.856&1.051&$36.2\pm6.5\pm4.1\pm5.7$&$0.73\pm0.14\pm0.10$\\
\hline\hline
\end{tabular}
\label{data}
 \ecl \end{table*}

The selection of charged tracks, photons, and $K_S^0\to \pp$
candidates are described in Refs.~\cite{ref13,guoyp}. A candidate
$\pi^0~(\eta)$ is reconstructed from a pair of photons with an
invariant mass in the range $|M_{\GG}-m_{\pi^0}|<$15~MeV/$c^2$
($|M_{\GG}-m_{\eta}|<$15~MeV/$c^2$), where $m_{\pi^0}~(m_{\eta})$ is
the nominal $\pi^0~(\eta)$ mass~\cite{ref14}.
The event candidates of $\EE\to \pphc$, $h_c\to\gamma\eta_c$ are required to have at least one
$\gamma\pi^0\pi^0$ combination with the mass recoiling against
$\pi^0\pi^0$, $M_{\pi^0\pi^0}^{\rm recoil}$, in the $h_c$ mass region
($M_{\pi^0\pi^0}^{\rm recoil}\in [3.3,3.7]$~GeV/$c^2$) and with the mass
recoiling against $\gamma\pi^0\pi^0$, $M_{\gamma\pi^0\pi^0}^{\rm
recoil}$, in the $\eta_c$ mass region ( $M_{\gamma\pi^0\pi^0}^{\rm
recoil}\in [2.8,3.2]$~GeV/$c^2$).

To determine the species of final state particles and to select the
best photon candidates when additional photons (and $\pi^0$ or $\eta$ candidates)
are found in an event, the combination with the minimum value of
$\chi^2= \chi^2_{4C} + \Sigma^N_{i=1}\chi^2_{\rm PID}(i) + \chi^{2}_{1C}$
is selected for further analysis. Here $\chi^2_{4C}$ is the $\chi^2$
of the initial-final four-momentum conservation ($4C$) kinematic
fit, $\chi^2_{\rm PID}(i)$ is the $\chi^2$ 
of particle identification (PID) of each charged track using the energy loss in the main
drift chamber and the time measured with the time-of-flight system,
$N$ is the number of the charged tracks, and $\chi^{2}_{1C}$ is the
sum of the $1C$ $\chi^2$s of the $\pi^0$s and $\eta$ in each final
state with the invariant mass of the daughter photon pair constrained to that of the
parent. There is also a $\chi^2_{4C}$ requirement, which is optimized
by maximizing the figure of merit $S/\sqrt{S+B}$, where $S$ and $B$ are the
numbers of Monte Carlo (MC) simulated signal and background events,
respectively.
The requirement $\chi^{2}_{4C} < 30$ has an efficiency of 82\% for
$\eta_c$ decays with only charged or $K^0_S$ particles in the final
states, while the requirement $\chi^{2}_{4C} < 25$ has an efficiency of
81\% for the other decays~\cite{corr}. A similar
optimization is performed to determine the $\eta_c$ candidate mass window
around its nominal value, 
which is found to be $\pm 35$~MeV/$c^2$. This mass window contains 77\% of
$\eta_c$ decays with only charged or $K^0_S$
particles in final states and 74\% for the other decays.

\begin{figure}[hbtp]
\centering
\includegraphics[width=0.5\textwidth]{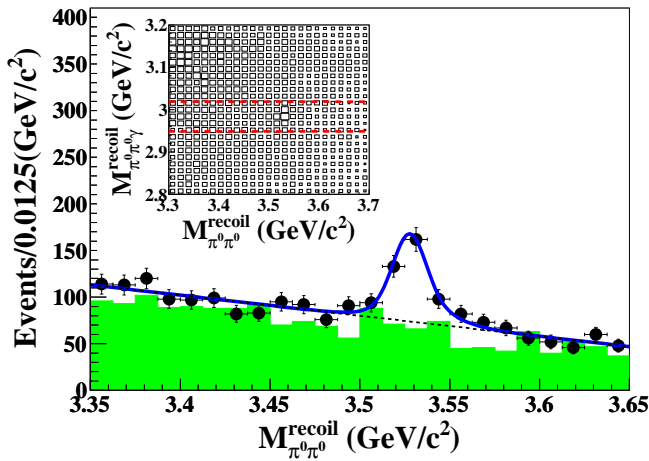}
\caption{The $M^{\rm recoil}_{\pi^0\pi^0}$ distribution for the events
with an $\eta_{c}$ candidate. The plot shows the sum over three CME
points. Dots with error bars are data; the solid curve is the best
fit; the dashed black line is the background; the green shaded histogram shows the normalized $\eta_c$ sideband
events. The inset shows the scatter plot of $M_{\gamma\pi^0\pi^0}^{\rm recoil}$ versus
$M_{\pi^0\pi^0}^{\rm recoil}$. The two red dashed lines represent the signal region of $\eta_c$.}
\label{scatter}
\end{figure}

The inset of Fig.~\ref{scatter} shows the scatter plot of
$M_{\gamma\pi^0\pi^0}^{\rm recoil}$,
which corresponds to the invariant mass of the
reconstructed $\eta_c$ candidate, versus $M_{\pi^0\pi^0}^{\rm recoil}$,
which corresponds to the
invariant mass of the reconstructed $h_c$ candidate, summed over the events at
$\sqrt{s}=4.23$, 4.26, and 4.36~GeV, where a clear cluster of events
corresponding to the $h_c\to \gamma\eta_c$ signal is observed.
Figure~\ref{scatter} shows the projection of the invariant mass
distribution of $\gamma \eta_c$ candidates for events in the $\eta_c$
signal region ($M_{\gamma\pi^0\pi^0}^{\rm recoil}\in
[2.945,3.015]$~GeV/$c^2$ ),
where a clear peak at the $h_c$ mass is
observed. The events in the sideband regions, 2.865 GeV/c$^2$ $<
M_{\gamma\pi^0\pi^0}^{\rm recoil} <$ 2.900 GeV/c$^2$ and
3.050 GeV/c$^2$$< M_{\gamma\pi^0\pi^0}^{\rm recoil} <$ 3.085 GeV/c$^2$
are used to study the background. To extract the number of $\pphc$
signal events, the $M^{\rm recoil}_{\pi^{0}\pi^{0}}$ mass spectrum is
fitted with a MC simulated signal shape convolved with a Gaussian
function to represent the data-MC mass resolution difference,
together with a linear background term.
A simultaneous fit to the $M^{\rm recoil}_{\pi^{0}\pi^{0}}$ mass
spectrum summed over the 16 $\eta_c$ decay modes at the three CME
points yields the numbers of $\pphc$ signal events
($n_{h_c}^{\rm obs}$) listed in Table~\ref{data}. Figure~\ref{scatter} also shows the fit results
summed over the three CME points.

The Born cross section $\sigma^{\rm B}(\EE \to\pi^0\pi^0 h_c)$ is calculated with the formula
\begin{align}
\sigma^{\rm B}&(e^{+}e^{-}\to\pi^{0}\pi^{0}h_{c}) = \nonumber \\
\quad & \textstyle \frac{n^{\rm
    obs}_{h_{c}}}{{\cal
    L}(1+\delta^{r})(1+\delta^{v})\sum\limits^{16}_{i=1}\epsilon_i{\cal
    B}(\eta_c\to X_{i}){\cal B}(h_c\to\gamma\eta_c)},
\label{eq:born_xs}
\end{align}
where $n^{\rm obs}_{h_{c}}$ is the number of observed $h_{c}$ signal
events; ${\cal L}$ is the integrated luminosity; $(1+\delta^{r})$ is
the initial radiative correction factor, which is taken to be the same
as that for the analysis of $\EE\to \pp h_c$~\cite{guoyp};
$(1+\delta^{v})$ is the vacuum polarization factor~\cite{vacu};
$\epsilon_i$ is the detection efficiency for the {\it $i^{th}$}
$\eta_c$ decay mode in the decay $\EE\to\pi^{0}\pi^0 h_c$ without
consideration of any possible intermediate structures and with ISR and
vacuum polarization effects considered in the MC simulation; ${\cal
  B}(\eta_c\ar X_i)$ is the corresponding $\eta_c$ branching fraction;
${\cal B}(h_c\to\gamma\eta_c)$ is the branching fraction of
$h_c\ar\gamma\eta_c$.

The measured Born cross sections are listed in
Table~\ref{data}. The ratios of the Born
cross sections for the neutral and charged $\EE\to \pi\pi h_c$ modes
are also listed in Table~\ref{data};  the cross sections for the
charged channel are taken from Ref.~\cite{guoyp},
where vacuum polarization effects were not taken into account.
A corresponding correction factor $(1+\delta^v)$ is applied to the previous Born cross section.
The common systematic uncertainties in the two measurements cancel in the ratio calculation.
The combined ratio ${\cal R}_{\pi\pi h_c}$ is obtained with a weighted
least squares method~\cite{xx} and determined to be
$(0.63\pm 0.09)$, which is within 2$\sigma$ of the expectation of isospin symmetry, 0.5.

Systematic uncertainties in the cross section measurement mainly come
from the luminosity measurement ($\delta_{\cal L}$), branching fraction of $h_c\to
\gamma\eta_c$, branching fractions of $\eta_c\to X_i$,
detection efficiencies ($\delta_{\epsilon_i\cdot{\cal B}(\eta_c\ar X_i)}$),
radiative correction factors ($\delta_{\rm ISR}$), vacuum polarization
factors ($\delta_{\rm Vac}$)~\cite{vacu}, and fits to
the mass spectrum.  The integrated luminosity at each CME points is
measured using large-angle Bhabha events and has an estimated
uncertainty of 1.0\%. The $h_c\to \gamma\eta_c$
and $\eta_c\to X_i$ branching fractions are taken from Refs.~\cite{ref11, ref13}, and the
uncertainties in the radiative correction are the same as those used in the analysis of
$\EE\to \pp h_c$~\cite{guoyp}. The uncertainties in the vacuum
polarization factor are 0.5\%~\cite{vacu}. The detection efficiency
uncertainty estimates are done with the same way as described in
Refs.~\cite{ref13,ref16}.
The uncertainty due to the $\eta_c$ mass ($\delta_{\eta_c-\rm mass}$) is
estimated by changing its mass by $\pm$1$\sigma$ of its world average value~\cite{ref14};
the uncertainties due to the background shapes ($\delta_{\rm bkg}$) are estimated
by changing the background function from a first-order to a second-order polynomial;
the uncertainty from the mass resolution ($\delta_{\rm res}$) is
estimated by varying the mass resolution difference between data and MC simulation by
one standard deviation; the uncertainty from fit range ($\delta_{\rm fit}$) is
estimated by extending the fit range; the uncertainty from the
$\pi^{0}\pi^{0}h_{c}$ substructure ($\delta_{\rm sub}$) is estimated
by considering the efficiency with and without the inclusion of a
$Z_{c}(4020)^{0}$. The contribution from each source of systematic
error are listed in Table~\ref{II}.

Assuming all of the above uncertainties are independent, the total
systematic uncertainties in the $\EE\to \pphc$ cross section
measurements are determined to be between 10\% and 13\%.  The uncertainty in ${\cal B}(h_c\to
\gamma\eta_c)$, not listed in Table~\ref{II} but common to all CME
points, is 15.7\%~\cite{ref14} and is quoted separately in the cross section measurement.

\begin{table*}[htbp]
\caption{The systematic uncertainties (\%) in $\sigma^{\rm B}(\EE\to \pphc)$.}
 \bcl \doublerulesep 2pt

\begin{tabular}{cccccccccc}\hline\hline
$\sqrt{s}$~(GeV)& $\delta_{\cal{L}}$ &$\delta_{\rm fit}$& $\delta_{\rm res}$
&$\delta_{\rm bkg}$&$\delta_{\eta_c-\rm mass}$  & $\delta_{\rm sub}$&$\delta_{\rm ISR}$&$\delta_{\rm Vac}$&$\delta_{\epsilon_i{\cal B}(\eta_c\ar X_i)}$\\\hline
4.230&1.0 &1.3&0.9&5.9&1.6&2.1&2.2&0.5&7.2\\
4.260&1.0 &0.9&0.4&9.5&4.8&1.6&2.3&0.5&7.3\\
4.360&1.0 &1.0&0.1&7.1&4.6&0.6&0.4&0.5&7.2\\
\hline\hline
\end{tabular}

\label{II}

 \ecl
  \end{table*}

\begin{figure}[htbp]
\centering
\includegraphics[width=0.5\textwidth]
{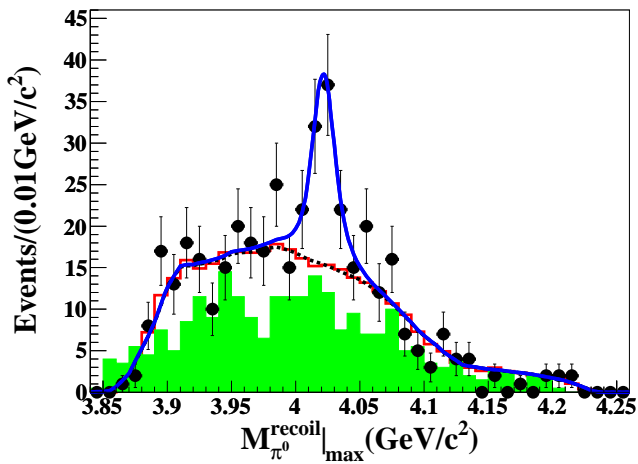}
\caption{Sum of the simultaneous fit to the $M^{\rm recoil}_{\pi^0}|_{\rm max}$ distribution at $\sqrt{s}=4.23$, 4.26 and
4.36~GeV as described in the text. Dots with errors bars are data;
the green shaded histogram shows the normalized $h_c$ sideband
events; the black dashed curve is the background from the fit; the
red histogram shows the result from a phase space MC simulation.
The solid blue line shows the total fit.}
\label{fitzc}
\end{figure}

Intermediate states are studied by examining the $M^{\rm
recoil}_{\pi^0}$ distribution (which corresponds to the
reconstructed $\pi^0 h_c$ invariant mass) for the selected
$\pphc$ candidate events. The $h_c$ signal events are selected by requiring 
$M_{\pi^0\pi^0}^{\rm recoil}$ in the range of [3.51, 3.55],
and events in the sideband regions [3.45, 3.49] and [3.57, 3.61]
 are used to study the background. From the two
combinations of the $\pi^{0}$ recoil mass in each event, we retain the
one with the larger $\pi^{0}$ recoil mass value, and denote this as
$M^{\rm recoil}_{\pi^0}|_{\rm max}$.
Figure~\ref{fitzc} shows the
$M^{\rm recoil}_{\pi^0}|_{\rm max}$ distribution for the signal events
where there is an obvious peak near 4.02~GeV/$c^2$,
which corresponds to the expected position of a $\zcpn$ signal.


An unbinned maximum likelihood fit is applied to the $M^{\rm recoil}_{\pi^0}|_{\rm max}$
distribution summed over all 16 $\eta_c$ decay modes. The data at
$\sqrt{s}=4.23$, 4.26, and 4.36~GeV are fitted simultaneously with
the same signal function with common mass and width.
The signal shape is parametrized with a constant-width
relativistic Breit-Wigner function convolved with a Gaussian-distributed
mass resolution, where
the mass resolution is determined from a fit to a MC sample with the
width set to zero. Because of the limited statistics of the $\zcpn$
signal,
its width is fixed to that of its charged partner, ($7.9\pm2.6$) MeV~\cite{guoyp}.
Assuming the spin and parity of the $\zcpn$ are
$1^+$, a phase space factor $pq^3$ is included in the partial width,
where $p$ is the $\zcpn$ momentum in the $\EE$ rest frame and $q$ is
the $h_c$ momentum in the $\zcpn$ rest frame.

There are two types of backgrounds in the $M^{\rm recoil}_{\pi^0}|_{\rm max}$ distribution.
One is the non-$h_c$ background in the $h_c$ signal region, which can
be represented by the $h_c$ sideband events, and the other is the
non-$\zcpn$ $\pphc$ events that may come from three-body $\pphc$ decays or
from production of intermediate scalar states, such as the $f_0(980)$,
that decay into $\pi^0\pi^0$. Since the widths of the low-mass scalar
particles  are large, these non-$\zcpn$ $\pphc$ events can be
reasonably well described with a phase space distribution.
For the non-$h_c$ background, a comparison of the $h_c$ sideband
events with the simulated phase space events indicates that it can
also be described with a three-body phase space distribution. Thus, in
the fit all of the background sources are described with a single
MC-simulated phase space shape with a total normalization that is left
as a free parameter.
In the fit, the signal shape mentioned above is multiplied
by the efficiency, which depends on $M^{\rm recoil}_{\pi^0}|_{\rm max}$.
Interference between the signal and background is neglected.

The solid curve in Fig.~\ref{fitzc} shows the fit results, which
yields a $\zcpn$ mass of $4023.9\pm 2.2$~MeV/$c^2$. The mass
difference between neutral and charged $\zcp$ is
$1.0\pm2.3$ (stat.)~MeV/$c^2$, which agrees with zero within error.
By projecting the events into a histogram with 50 bins, the goodness of the fit is
calculated from the combined $\chi^{2}$ values, the number of bins and
the number of free parameters at three CME points, and found to be
$\chi^2/n.d.f.=28.6/33$. Here the event number in each bin used in the
$\chi^2$ evaluation is required to be larger than 7.  The statistical
significance of the $\zcpn$ signal is determined from a comparison of
the fit likelihoods with and without the signal. Additional fit are
also performed with different signal shapes, and background shapes. In
all cases, the minimum significance is found to be above
$5\sigma$. The numbers of $\zcpn$ signal events are listed in
Table~\ref{III}.

The Born cross section $\sigma^{\rm B}(\EE\to \pi^{0}\zcpn\to\pi^0\pi^0 h_c)$ is
calculated with eq.~\ref{eq:born_xs}, with the measured numbers of
observed signal and MC-determined detection efficiencies for the
$\pi^0Z_c(4020)^0$ channel.

\begin{table*}[htbp]
 \caption{Energies ($\sqrt{s}$), numbers of events ($n_{\zcpn}^{\rm
     obs}$), initial state radiative correction factor ($1+\delta^{r}$)~\cite{guoyp}, vacuum polarization factor ($1+\delta^v$), average efficiencies
   ($\sum\limits^{16}_{i=1}\epsilon_i{\cal B}(\eta_c\ar X_i)$),
    Born cross sections $\sigma(\EE\to \pi^{0}
 \zcpn\to\pi^0\pi^0 h_c)$, and ratios ${\cal R}_{\pi Z_c(4020)}=\frac{\sigma(\EE\to
 \pi^{0}\zcpn\to\pi^{0}\pi^{0}h_{c})} {\sigma(\EE\to \pi^{\pm}\zcp^{\mp}\to\pi^{\pm}\pi^{\mp}h_{c})}$, where the third
 errors are from the uncertainty in ${\cal B}(h_c\to
 \gamma\eta_c)$~\cite{ref13}.}  \bcl \doublerulesep 2pt
\begin{tabular}{ccccccc}\hline\hline
 $\sqrt{s}~({\rm GeV})$ & $n_{\zcpn}^{\rm obs}$&$(1+\delta^r)$
  &$1+\delta^{v}$ &$\sum\limits^{16}_{i=1}\epsilon_i{\cal B}(\eta_c\ar X_i)$ &$\sigma^{\rm B}(\EE\to \pi^{0} \zcpn\to\pi^0\pi^0 h_{c})~({\rm pb})$ & $R_{\pi Z_{c}(4020)}$\\\hline
4.230&$21.7\pm7.4$&0.756&1.056&$7.08\times10^{-3}$ &$6.5\pm2.2\pm0.7\pm1.0$&$0.77\pm0.31\pm0.25$\\
4.260&$22.5\pm7.7$&0.831&1.054&$6.72\times10^{-3}$ &$8.5\pm2.9\pm1.1\pm1.3$&$1.21\pm0.50\pm0.38$\\
4.360&$17.2\pm7.2$&0.856&1.051&$6.56\times10^{-3}$ &$9.9\pm4.1\pm1.3\pm1.5$&$1.00\pm0.48\pm0.32$\\
\hline\hline
\end{tabular}
\label{III}
 \ecl \end{table*}

\begin{table*}[ht]
\caption{Systematic uncertainties~(\%) in the $\sigma(\EE\to
\pi^0\zcpn\to \pphc$) measurement, in addition to the common part of those in
$\sigma(\EE\to \pphc$).}
 \bcl \doublerulesep 2pt
\begin{tabular}{ccccccc}\hline\hline
$\sqrt{s}$~(GeV)& $\delta_{\rm signal}$ & $\delta_{\rm bkg}$&$\delta_{\rm res}$&$\delta_{ h_{c}-\rm signal}$&$\delta_{\epsilon_{\rm curve}}$&$\delta_{\rm MC-model}$\\\hline
4.230&0.3 &5.8&0.5&5.1&0.3&0.6\\
4.260&1.1 &3.5&0.2&8.6&0.3&0.6\\
4.360&0.8 &4.8&0.2&3.5&0.0&0.6\\
\hline\hline
\end{tabular}
\label{IV}
 \ecl \end{table*}

The systematic uncertainties on the $\zcpn$ mass
come from uncertainties in the mass
calibration and energy scale, parametrizations of the signal and background shapes,
mass dependence of the efficiency, width assumption, MC modeling with a different $J^P$ value, and mass
resolution. The uncertainty from the mass calibration is estimated by
using the difference, $(2.3\pm 1.5)$ MeV/c$^{2}$, between the measured and known $h_c$ mass. The
uncertainty from the photon energy scale is estimated with
$\psp\to\gamma\chi_{c1,2}, \chi_{c1,2}\to\gamma J/\psi,
J/\psi\to\mu^{+}\mu^{-}$ for photons with low energy, and with
radiative Bhabha processes for photons with high
energy~\cite{ref11}. After adjusting the MC energy scale accordingly,
the resulting changes in the mass of $\zcpn$ are negligible. The $J^P$
value of $\zcpn$ is uncertain; two possible alternatives,
$J^P=1^-$ and $2^+$, are used to estimate the corresponding systematic errors. A
difference of 0.4~MeV/$c^2$ in the $\zcpn$ mass is found under
different $J^P$ assumptions. The uncertainty due to the background shape is
determined by changing the phase space shape to a parametrized
background function, $f(M)=[(M-M_a)^{1/2}+c_1(M-M_a)^{3/2}]\times[(M_b-M)^{1/2}+c_2(M_b-M)^{3/2}]$. Here $M$ is mass of the background, $M_a$ and $M_b$ are the two extreme points determined by the minimal and maximal mass. $f(M)=0$ for $(M-M_a)<0$ or $(M_b-M)<0$. The coefficients $c_1$ and $c_2$ are determined by the fit~\cite{ref5add}.
A difference of 0.1~MeV/$c^2$ is
found and taken as the systematic uncertainty. The uncertainty due to
the mass dependence of the efficiency is determined by assuming a
uniform efficiency in the whole $M^{\rm recoil}_{\pi^0}|_{\rm max}$
recoil mass region, and the difference is found to be negligible.
The uncertainty due to the
mass resolution is estimated by varying the data-MC difference in
resolution by one standard deviation of the measured uncertainty in
the mass resolution of the $h_c$ signal; the difference in the
$Z_c(4020)^0$  mass is negligible. Similarly, the uncertainty due
to the fixed $\zcpn$ width is estimated by varying the width determined
for its charged partner by one standard deviation. The difference is
0.1~MeV/$c^2$ and is taken as the systematic error. Assuming all the
sources of the systematic uncertainty are independent, the total
systematic error is estimated to be 3.8~MeV/$c^2$.

The systematic uncertainties in the measured Born cross section,
$\sigma(\EE\to \pi^0\zcpn \to \pphc)$, are estimated in the same way
as for $\EE\to \pphc$. In addition to those common parts in the
$\EE\to \pphc$ measurement, the uncertainties due to signal parametrization ($\delta_{\rm signal}$),
background shape ($\delta_{\rm bkg}$), $h_c$ signal window selection ($\delta_{h_c-\rm{signal}}$),
mass resolution ($\delta_{\rm res}$),
efficiency ($\delta_{\epsilon_{\rm curve}}$), and MC model
($\delta_{\rm MC-model}$) are considered;
their values are summarized in Table~\ref{IV}.

The ratios of Born cross section for $\EE\to \pi Z_c(4020)\to\pi\pi h_c$ between
neutral and charged modes at three center-of-mass energies are listed
in Table~\ref{III}. Similar to the calculation of the
$\sigma(\EE\ar\pphc)$ ratio, the same correction factor $(1+\delta^v)$
is also applied to the previously measured
$\EE\ar\pi^{\pm}Z_c(4020)^{\mp}$ Born cross section. The common
systematic uncertainty between neutral and charged mode cancel.
The combined ratio ${\cal R}_{\pi Z_c(4020)}$ is determined to be
$(0.99\pm 0.31)$ with the same method as for the combined ${\cal
 R}_{\pi\pi h_c}$, which is well within 1$\sigma$ of the expectation of
isospin symmetry, 1.0.

In summary, we observe $\EE\to \pphc$ at $\sqrt{s}=4.23$, 4.26, and
4.36~GeV for the first time. The measured Born cross sections are about half of those for
$\EE\to \pi^+\pi^-h_c$, and agree with expectations based on isospin symmetry within systematic uncertainties.
A narrow structure with a mass of $(4023.9\pm
2.2\pm 3.8)$~MeV/$c^2$ is observed in the $M^{\rm recoil}_{\pi^0}|_{\rm max}$ mass
spectrum. This structure is most likely the neutral isospin partner of the
charged $Z_c(4020)$ observed in the $\EE\to\pi^{+}\pi^{-}h_c$ process~\cite{guoyp}.
This observations indicate that there is no anomalously large isospin violations in $\pi\pi h_c$ and $\pi Z_c(4020)$ system.

The BESIII collaboration thanks the staff of BEPCII and the IHEP computing
center for their strong support. This work is supported in part by National
Key Basic Research Program of China under Contract No. 2015CB856700;
Joint Funds of the National Natural Science Foundation of China under Contracts
Nos. 11079008, 11179007, U1232201, U1332201; National Natural Science
Foundation of China (NSFC) under Contracts Nos. 10935007, 11121092,
11125525, 11235011, 11079023, 11322544, 11335008; the Chinese Academy
of Sciences (CAS) Large-Scale Scientific Facility Program; CAS under
Contracts Nos. KJCX2-YW-N29, KJCX2-YW-N45; 100 Talents Program of CAS;
German Research Foundation DFG under Contract No. Collaborative Research
Center CRC-1044; Istituto Nazionale di Fisica Nucleare, Italy; Ministry
of Development of Turkey under Contract No. DPT2006K-120470; Russian
Foundation for Basic Research under Contract No. 14-07-91152;
U. S. Department of Energy under Contracts Nos. DE-FG02-04ER41291,
DE-FG02-05ER41374, DE-FG02-94ER40823, DESC0010118; U.S. National
Science Foundation; University of Groningen (RuG) and
the Helmholtzzentrum fuer Schwerionenforschung GmbH (GSI), Darmstadt;
WCU Program of National Research Foundation of Korea under Contract
No. R32-2008-000-10155-0.

\end{document}

%% file: def-com.tex
\newcommand{\ppp}{\pi^+\pi^- \pi^0}

\newcommand{\jp}{J/\psi\rightarrow \gamma\pi^0}

\newcommand{\EE}{e^+e^-}

\newcommand{\GG}{\gamma\gamma}

\newcommand{\pp}{\pi^+\pi^-}
\newcommand{\kk}{K^+K^-}
\newcommand{\ppb}{p\bar{p}}

\newcommand{\ppkk}{\pi^+\pi^-K^+K^-}

\newcommand{\psp}{\psi^{\prime}}
\newcommand{\jpsi}{J/\psi}
\newcommand{\ar}{\rightarrow}

\newcommand{\ppjpsi}{\pi^+\pi^- J/\psi}
\newcommand{\pphc}{\pi^0\pi^0h_c}

\newcommand{\bfg}{\begin{figure}}
\newcommand{\efg}{\end{figure}}
\newcommand{\bitm}{\begin{itemize}}
\newcommand{\eitm}{\end{itemize}}
\newcommand{\bnum}{\begin{enumerate}}
\newcommand{\enum}{\end{enumerate}}
\newcommand{\btbl}{\begin{table}}
\newcommand{\etbl}{\end{table}}
\newcommand{\btbu}{\begin{tabular}}
\newcommand{\etbu}{\end{tabular}}
\newcommand{\bcl}{\begin{center}}
\newcommand{\ecl}{\end{center}}

\newcommand{\beq}{\begin{equation}}
\newcommand{\eeq}{\end{equation}}
\newcommand{\beqr}{\begin{eqnarray}}
\newcommand{\eeqr}{\end{eqnarray}}